\documentstyle[12pt]{article}
\def\a{\alpha'}
\def\R{{\mathcal R}}
\def\T{{\mathcal T}}
\newcommand{\be}{\begin{equation}}
\newcommand{\ee}{\end{equation}}
\newcommand{\bea}{\begin{eqnarray}}
\newcommand{\eea}{\end{eqnarray}}
\begin{document}

\begin{center}

\vskip 20mm

{\Huge Perturbative stability and absorption cross section in string
corrected black holes}

\vskip 10mm

Filipe Moura

\vskip 4mm

{\em Security and Quantum Information Group - IT\\Instituto Superior
T\'ecnico, Departamento de Matem\'atica \\Av. Rovisco Pais, 1049-001
Lisboa, Portugal}

\vskip 4mm

{\tt fmoura@math.ist.utl.pt}

\vskip 6mm
\end{center}
\vskip .2in

\begin{center} {\bf Abstract } \end{center}
\begin{quotation}
\noindent We consider a $d-$dimensional spherically symmetric
dilatonic $\R^2$ string corrected black hole solution. We study its
stability under tensor type gravitational perturbations and compute
the absorption cross section for low frequency gravitational waves.
\end{quotation}

\vskip 10mm

We are interested in studying the behavior of a string-corrected
dilatonic black hole solution under perturbations in $d$ spacetime
dimensions \cite{kodama-ishibashi-0}, setting any tensorial or
fermionic fields to zero and taking as background metric \be
d\,s^2=-f(r)\,d\,t^2 + f^{-1}(r)\,d\,r^2 +r^2\,d\,\Omega^2_{d-2}
\label{schwarz} \ee with $d\Omega_{d-2}^2=\gamma_{ij}
\left(\theta\right)\,d \theta^i\,d \theta^j$,
$\gamma_{ij}=g_{ij}/r^2$ being the metric of a $(d-2)$-sphere
$S^{d-2}$.

We take, in the effective action, only the leading $\R^2$ $\a$
correction \cite{cmp89}: \be \frac{1}{2 \kappa^2} \int \sqrt{-g}
\left[\R-\frac{4}{d-2} \left(\partial^\mu \phi \right)
\partial_\mu \phi + \mbox{e}^{\frac{4}{2-d} \phi}
\frac{\lambda}{2} \R^{\mu \nu \rho \sigma} \R_{\mu \nu \rho \sigma}
\right] \mbox{d}^dx, \label{eef} \ee with $\lambda=\frac{\a}{2},
\frac{\a}{4}, 0$ for bosonic, heterotic and superstrings,
respectively.

The corrected field equation for the graviton is, to this order, \be
\R_{\mu \nu}+ \lambda \mbox{e}^{\frac{4}{2-d} \phi} \left(\R_{\mu
\rho \sigma \tau} \R_\nu^{\ \rho \sigma \tau} - \frac{1}{2(d-2)}
g_{\mu \nu} \R_{\rho \sigma \lambda \tau} \R^{\rho \sigma \lambda
\tau} \right) = 0. \label{bgfe} \ee

Here we only take tensorial perturbations to the metric (by using
the dilaton field equation, we show that we can set $\delta
\phi=0$), given by $h_{\mu\nu}=\delta g_{\mu\nu}$
\cite{kodama-ishibashi-2}: \be h_{ij}=2 r^2 H_T(r)
\T_{ij}\left(\theta^i \right), \, h_{ir}= h_{it}=0, h_{rr}=
h_{tr}=h_{tt}=0. \label{htensor} \ee $D_i$ is the $S^{d-2}$
covariant derivative; $\T_{ij}$ are the eigentensors of the
$S^{d-2}$ laplacian, with eigenvalues $-k_T=2-\ell \left(\ell
+d-3\right), \ell=2, 3, 4\ldots,$ satisfying \be \left(\gamma^{kl}
D_k D_l + k_T \right) \T_{ij}= 0, \, D^i \T_{ij}=0, \, g^{ij}
\T_{ij}=0. \label{propt} \ee

Using the explicit form of the Riemann tensor for the metric
(\ref{schwarz}) and its variations, computed from (\ref{htensor}),
and perturbing (\ref{bgfe}), we determine the equation for $H_T,$
which we write in the form of a master equation (with $r_*$ defined
by $d r_*/d r=1/f$) \be \frac{\partial^2 \Phi}{\partial r_*^2} -
\frac{\partial^2 \Phi}{\partial t^2} =: V_T \Phi, \label{ikmaster}
\ee As explained in \cite{Moura:2006pz}, we derive our master
function and potential: \bea \Phi&=& \frac{H_T}{\sqrt{f}}
\exp\left(\int \frac{\frac{f'}{f} +\frac{d-2}{r} + \frac{4}{r^3}
(d-4) \lambda (1-f) - \frac{4}{r^2} \lambda f' - \frac{2}{r f}
\lambda f'^2}{2-\frac{4}{r}\lambda f'} dr\right), \nonumber \\
V_{\textsf{T}} [f(r)]&=& \left(\frac{1}{1-2 \lambda
\frac{f'}{r}}\right)^2 \left(1+ \frac{4 \lambda}{r^2} \left(1-f
\right) \right) \left[ \frac{d-4}{4r^2} \left(1+ \frac{4
\lambda}{r^2} \left(1-f \right) \right) + \frac{2 \lambda f''
-1}{2r^2} \right] \nonumber \\ &+& \frac{1}{1-2 \lambda
\frac{f'}{r}} \left[\left(k_T +2\right) \frac{f}{r^2} + 2 (d-3)
\frac{f (1-f)}{r^2} + \frac{d-8}{2} \frac{f f'}{r} -
\frac{\lambda}{d-2} f \left( f''\right)^2 \right. \nonumber \\ &+&
\left. 4 \lambda  \left(k_T + 2 \right) \frac{f (1-f)}{r^4} + 2(d-3)
\lambda \frac{f (1-f)^2}{r^4}+ 2 (d-4) \lambda \frac{f(1-f) f'}{r^3}
\right] \nonumber \\ &+& \frac{f f'}{r} + (d-4) \frac{f^2}{r^2}.
\eea
 To study the stability of a solution, we use the
``S-deformation approach'' \cite{kodama-ishibashi-2}. After having
obtained the potential $V_T (f)$, we assume that its solutions are
of the form $\Phi(r_*,t)=e^{i\omega t} \phi(r_*),$ such that
$\partial\Phi/\partial t= i\omega \Phi.$ The master equation is then
written in the Schr\"odinger form $A \Phi= \omega^2 \Phi$, and a
solution to the field equation is then stable if the operator $A$ is
positive definite with respect to the following inner product: \be
\left\langle \phi, A \phi \right \rangle = \int_{-\infty}^{+\infty}
\overline{\phi} (r_*) \left[ -\frac{d^2}{dr_*^2} +V \right]
\phi(r_*) \ dr_* = \int_{-\infty}^{+\infty} \left( \left|D \phi
\right|^2 + \frac{Q}{f} \left|\phi \right|^2 \right) \ dr_*\ee (see
\cite{Moura:2006pz} for the details), with $D=\frac{d}{dr_*}-\frac{f
H_T}{\Phi} \frac{d}{dr} \left( \frac{\Phi}{H_T}\right)$ and \bea
\frac{Q}{f}&=& \frac{1}{1-2 \lambda \frac{f'}{r}} \frac{1}{r^2}
\left[\left(k_T +2\right) \left(1+ \frac{4 \lambda}{r^2} \left(1-f
\right) \right) + (2d-6) (1-f) \left(1+ \frac{\lambda}{r^2}
\left(1-f \right) \right) \right. \nonumber \\ &-& \left. 2r f' -
\frac{\lambda}{d-2} \left( f''\right)^2 r^2 \right]. \label{qf} \eea
All that is necessary to guarantee the stability is to check the
positivity of $\frac{Q}{f}.$

In \cite{Moura:2006pz} we considered the $\R^2$-corrected black hole
solution of the type of (\ref{schwarz}) studied in \cite{cmp89},
taking a coordinate system in which the horizon radius $R_H$ is not
changed. Assuming $R_H \gg \sqrt{\lambda}$, for this solution $f(r)$
is given by \be f(r)=\left(1-\left(\frac{R_H}{r}\right)^{d-3}
\right) \left[1- \lambda \frac{(d-3)(d-4)}{2}
\frac{R^{d-5}_H}{r^{d-1}} \frac{r^{d-1}-R_H^{d-1}}{r^{d-3} -
R^{d-3}_H} \right]. \label{fr2} \ee We showed\cite{Moura:2006pz}
that $\frac{Q}{f}>0$; therefore this solution is stable under tensor
perturbations.

In Einstein-Hilbert gravity, for any spherically symmetric black
hole in arbitrary dimension the absorption cross section of
minimally coupled massless scalar fields equals the area of the
black hole horizon \cite{dgm96}, a result which suggests a
universality of the low-frequency absorption cross sections of
generic black holes. Since the equation describing gravitational
perturbations to a black hole solution allows for a study of
scattering in this spacetime geometry, we tried to extend this
result to the higher-derivative corrected black hole (\ref{fr2}),
focusing only on the leading contribution to the scattering process:
the s-wave, with $\ell=0.$ The low-frequency regime $R_H \omega \ll
1$ allows us to fully analytically solve the problem by using the
technique of matching solutions near the event horizon to solutions
at asymptotic infinity. In both these regions the potential
$V_{\textsf{T}} [f(r)]$ vanishes, and the master equation reduces to
a free-field equation whose solutions are plane waves in the
tortoise coordinate.

Near the event horizon, $r \simeq R_H$, since we are computing the
absorption cross section, we shall consider the general solution for
an incoming plane wave $H_T(r_*) = A_{\mathrm{near}} e^{i \omega
r_*};$ after expanding $V_{\textsf{T}}(r)$ and $r_*(r)$ this
solution becomes \be H_T (r) \simeq A_{\mathrm{near}} \left( 1 +
i\frac{R_H \omega}{d-3} \left( 1 + \frac{(d-1) (d-4)}{2}\
\frac{\lambda}{R_H^2} \right) \log \left( \frac{r-R_H}{R_H} \right)
\right).\ee

Close to infinity, one must consider a superposition of incoming and
outgoing waves which becomes, expressed in the original radial
coordinate, $$H_T (r) = \left( r \omega \right)^{(3-d)/2} \left[
A_{\mathrm{as}}\, J_{(d-3)/2} \left( r\omega \right) +
B_{\mathrm{as}}\, N_{(d-3)/2} \left( r\omega \right) \right];$$ at
low frequencies, $r\omega \ll 1$, \be H_T (r) \simeq
A_{\mathrm{as}}\ \frac{1}{2^{\frac{d-3}{2}} \Gamma \left(
\frac{d-1}{2} \right)} + B_{\mathrm{as}}\ \frac{2^{\frac{d-3}{2}}
\Gamma \left( \frac{d-3}{2} \right)}{\pi \left( r\omega
\right)^{d-3}} + {\mathcal{O}} \left( r\omega \right).\ee In order
to compute the absorption cross section, one needs the fluxes per
unit area $J = \frac{1}{2i} \left( H_T^\dagger (r_*) \frac{d
H_T}{dr_*} - H_T (r_*) \frac{d H_T^\dagger}{dr_*} \right).$ Near the
horizon this quantity is given by $J_{\mathrm{near}} = \omega \left|
A_{\mathrm{near}} \right|^2;$ close to infinity we analogously have
$J_{\mathrm{as}} =\frac{2}{\pi} r^{2-d} \omega^{3-d} \left|
A_{\mathrm{as}} B_{\mathrm{as}} \right|.$ In order to match the
coefficients $A_{\mathrm{near}}$, $A_{\mathrm{as}}$ and
$B_{\mathrm{as}},$ one needs to interpolate between the solutions
near the event horizon and at asymptotic infinity. This requires
solving the master equation in the intermediate region between the
horizon and infinity. The full computation can be found in
\cite{Moura:2006pz}, where it is shown that \bea A_{\mathrm{as}} &=&
2^{\frac{d-3}{2}}
\Gamma \left( \frac{d-1}{2} \right) A_{\mathrm{near}}, \nonumber \\
B_{\mathrm{as}} &=& - \frac{i \pi \left( R_H \omega
\right)^{d-2}}{2^{\frac{d-3}{2}} (d-3) \Gamma \left( \frac{d-3}{2}
\right)} \left( 1 + \frac{(d-1) (d-4)}{2}\ \frac{\lambda}{R_H^2}
\right) A_{\mathrm{near}}. \eea With these results one obtains the
scattering cross section \be \sigma_T^{\ell=0}= \frac{\int r^{d-2}
J_{\mathrm{as}}d \Omega_{d-2}}{J_{\mathrm{near}}} = A_H \left( 1 +
\frac{(d-1) (d-4)}{2}\ \frac{\lambda}{R_H^2} \right).\ee

We conclude that the absorption cross section is increased due to
the $\a$ corrections, although it is still proportional to the area
of the event horizon. The same happens to the black hole entropy,
although its $\a$ correction has a different value
\cite{cmp89,Moura:2006pz}.

\paragraph{Acknowledgements}
\noindent Work supported by Funda\c c\~ao para a Ci\^encia e a
Tecnologia through Centro de L\'ogica e Computa\c c\~ao (CLC) and
fellowship BPD/14064/2003, in collaboration with Ricardo Schiappa.


\begin{thebibliography}{10}
\bibitem{kodama-ishibashi-0}
H.~Kodama, A.~Ishibashi and O.~Seto, \textit{Brane World Cosmology:
Gauge Invariant Formalism For Perturbation}, Phys.\ Rev.\
\textbf{D62} (2000) 064022, \texttt{[hep-th/0004160]}.
\bibitem{cmp89} C.~G.~Callan, R.~C.~Myers and M.~J.~Perry,
\textit{Black Holes in String Theory}, Nucl.\ Phys.\ \textbf{B311}
(1989) 673.
\bibitem{kodama-ishibashi-2}
A.~Ishibashi and H.~Kodama, \textit{Stability of Higher Dimensional
Schwarzschild Black Holes}, Prog.\ Theor.\ Phys.\ \textbf{110}
(2003) 901, \texttt{[hep-th/0305185]}.
\bibitem{Moura:2006pz}
F.~Moura and R.~Schiappa, \textit{Higher-derivative corrected black
holes: Perturbative stability and absorption cross-section in
heterotic string theory}, Class.\ Quant.\ Grav.\ {\bf 24} (2007)
361, \texttt{[hep-th/0605001]}.
\bibitem{dgm96}
S.~R.~Das, G.~Gibbons and S.~D.~Mathur, \textit{Universality of Low
Energy Absorption Cross--Sections for Black Holes}, Phys.\ Rev.\
Lett.\ \textbf{78} (1997) 417, \texttt{[hep-th/9609052]}.
\end{thebibliography}
\end{document}